\documentclass[doublespacing]{elsart}
\usepackage{stmaryrd}
\usepackage{amsmath}
\usepackage{amssymb}
\usepackage{amsfonts}
\usepackage{graphicx}
\usepackage{ctex}
\usepackage{rotating}
\usepackage{longtable}
\usepackage{graphics}
\usepackage{epstopdf}
\usepackage{epsfig}
\usepackage{color}\date{}
\begin{document}

\begin{frontmatter}



\title{Magnetism and electronic properties of BiFeO$_3$ under lower pressure}

\author{Hongjian  Feng}

\address{Department of Physics and Astronomy, University of
Missouri,Columbia,  MO 65211, USA} \ead{fenghongjian@gmail.com,
fengh@missouri.edu}

\begin{abstract}
\textcolor{red}{$Ab$ $initio$ calculations show  an
antiferromagnetic-ferromagnetic phase  transition around 9-10 GPa
and a magnetic anomaly at 12 GPa in BiFeO$_3$. The magnetic phase
transition also involves a structural and insulator-metal
transition. The G-type AFM configuration under pressure leads to an
increase of the $y$ component and decrease of the $z$ component of
the magnetization,  which is caused by the  splitting of the
$d_{z^2}$  orbital from doubly degenerate $e_g$ states. Our results
agree with recent experimental results.}

\end{abstract}

\end{frontmatter}


\section{ Introduction}
Multiferroic materials have more than two of
ferroelectric/antiferroelectric, ferromagnetic/ antiferromagnetic,
and even ferroelastic/ antiferroelastic ordering in same phase
\cite{1,2,3,4}. This feature makes them have potential applications
in information storage, spintronics,and sensors. The study on
BiFeO$_3$, a multiferroic,  which possess weak ferromagnetism and
ferroelectric properties simultaneously at room temperature, has
been recovered recently.\cite{5,6,7}. It has long been known to be
ferroelectric with a Curie temperature of about 1103 K and G-type
antiferromagnetic(AFM) with a N\'{e}el temperature of 643 K. The
large difference between the magnetic and ferroelectric ordering
temperature makes the linear magnetoelectric coefficient smaller.
Moreover,the $R3c$ space group permits a spiral spin structure in
which the AFM axis rotates through the crystal with a
 long-wavelength period of 620{\AA}, which further  reduces the observed  G-type AFM magnetization.
Fortunately, the decrease of magnetization can be compensated by
doping  in B-site of perovskites and fabricating thin film
samples\cite{8,9,10}. The ferroelectricity is produced by the Bi-6s
stereochemically  active lone pair , which can only occur if the
cation ionic site have broken inversion symmetry, while the weak
magnetism is mainly attributed to Fe$^{3+}$ ions. The coupling
between magnetic and ferroelectric parameter is weak because they
are driven by different ionic sites, and this agrees with the fact
of large difference between the Curie temperature  and AFM N\'{e}el
temperature. Through first-principles calculations, we have shown
that the rotation of the oxygen octahedra(antiferrodistortive(AFD)
distortion) couples with the weak ferromagnetism due to the
Dzyaloshinskii-Moriya interaction(DMI), considering the
spin-orbital(SO) coupling effect and the noncollinear spin
configuration\cite{11}.

The study of BiFeO$_3$ under pressure should give us much insight in
the understanding of the magnetoelectric coupling because it
involves interesting  structural and magnetic changes under
pressure. Another metal-insulator transition and magnetic anomaly
have already been reported recently\cite{12,13}. Meanwhile,  A. J.
Hatt and coworkers performed the first principles calculations on
strain induced phase transition of BiFeO$_3$ film on the
(001)-oriented substrate\cite{14,15}. An isosymmetric phase
transition accompanying with dramatic structural change has been
found. The two isosymmetric phases have the same space-group
symmetry due to the constraints caused by coherence and epitaxy.
The isosymmetric transition also indicates the coexistence of the
rhombohedral and tetragonal phases.  The rhombohedral film tends to
be expanded with large compressive strain. And the lattice expansion
can be accommodated by the increase of tetragonal domain. Further
theoretical calculations  about the phase transition of BiFeO$_3$
need to be done to  explain the detailed mechanism  under pressure.
We have performed first-principles calculations to investigate the
magnetic and electronic properties of BiFeO$_3$ under lower pressure
range corresponding to the experiments to shed light on the
mechanism. The remainder of this article is structured as follows:
In section 2, the computational details of our calculations are
given.  We discuss the results in section 3. Our conclusion are
given in section 4.

\section{ Computational details}

We performe calculations within the local spin density
 approximation(LSDA) to DFT using the
ABINIT package\cite{16,17}.  The ion-electron interaction is modeled
by the projector augmented wave (PAW) potentials \cite{18,19} with
energy cutoff of 500 eV. We treat Bi 5d, 6s, and 6p electrons, Fe
4s, 4p,and 3d electrons, and O 2s and 2p electrons
 as valence states. $10\times10\times10$
Monkhorst-Pack sampling of the Brillouin zone are used in
calculations.  LSDA+U method is introduced,  where the strong
coulomb repulsion between localized $d$ states has been considered
by adding a Hubbard-like term to the effective
potential\cite{20,21,22}.

\section{ Results and discussion}
\subsection{FM behavior}
According to  the experimental results and previous
calculations\cite{13,23}, we have constructed three structures,
rhombohedral, monoclinic, and orthorhombic structures, with space
group $R3c$, $Cm$, and $Pnma$, respectively. The lattice parameters
are given in Table 1. The phase transition under pressure have been
discussed elsewhere, and we suggest there exists a first-order phase
transition from $R3c$ to $Cm$ structure around 9-10 GPa, and
gradually to pure $Pnma$ structure around 12 GPa which is close to
the experimental values\cite{13}.We only report the magnetic and
electronic behavior in this work.  FM spin configuration for each
structure are used to analyze the magnetization under pressure. The
FM magnetization are shown in Fig.1. One can see the spin value for
the monoclinic structure is generally larger than the other two
structures, and it exhibits a decrease of spin magnetization with
increasing pressure and leads to a flat curve after passing 9-10
GPa.  The monoclinic and rhombohedral structures both have the same
spin value at all pressure range, indicating they have same
structure dependent spin configuration. Moreover the phase
transition between these two  can be much easier than  others. The
transition between these two structures happens  at 9-10 GPa. That
between rhombohedral and orthorhombic structures  take places at
higher pressure of 12 GPa.
\subsection{AFM-FM spin order transition}
 Meanwhile, for rhombohedral structure we set up three spin configurations, FM, AFM, and G-type AFM spin structures. The total energy per  unit cell for different spin configuration have been calculated and shown in Fig. 2,  as well as considering the on-site Coulomb interaction term, U.
In  LSDA+U calculations, $U$ and $J$ are defined as
\begin{equation}
U=\frac{1}{(2l+1)^2}\sum_{m,m'}<m,m'|V_{ee}|m,m'>=F^0,
\end{equation}
\begin{equation}
J=\frac{1}{2l(2l+1)}\sum_{m\neq
m',m'}<m,m'|V_{ee}|m,m'>=\frac{F^2+F^4}{14},
\end{equation}
where $V_{ee}$ are the screened Coulomb interaction among the $nl$
electrons. $F^0$, $F^2$, and$F^4$ are the radial Slater integrals
for $d$ electrons in Fe.  It is apparent that there are intersection
points between the FM and AFM energy curves. As pressure approaches
the intersection points, both LSDA+U and LSDA give lower AFM energy.
However FM energy are much favorable after pressure exceeds the
intersections, indicating an AFM-FM phase transition around the
pressure value of 9-10 GPa. The LSDA+U can produce  relatively
higher energy. On-site coulomb interaction , $U$ is the energy
needed to put two electrons in the same site. The value of $U$ among
the electrons in transition-metal $d$ orbitals are one magnitude
higher than the Stoner parameter. An appropriate band gap can be
obtained  in transition-metal oxides with properly choosing the $U$
parameter . In our calculation we choose $U$=2 eV as it is rightly
under the critical value to preclude the DMI in G-type AFM spin
configuration, where the Fe ions are arranged antiferromagnetically
along $x$ direction\cite{24}. We take into account the non-collinear
spin structure and spin-orbital(SO) interaction  in G-type
configuration. One can see that the G-type spin structure leads to a
lower energy comparing with the FM and AFM one within LSDA+U scheme.
Except the anomaly  around the first transition pressure of  9-10
GPa, there is another one around 12 GPa. The relaxation results show
the structure is changed into an orthorhombic phase at this
pressure.

\subsection{Exchange interaction}
Consideration the Heisenberg model,

\begin{equation}
\Delta E=-1/2\sum_{i,j}J_{i,j}\mathbf{S_i} \cdot \mathbf{S_j},
\end{equation}

for FM spin configuration,  the total energy involving the spin
exchange interaction can be written as,
\begin{equation}
E_{FM}=E_t-Z_cJ_zS^2,
\end{equation}
where $E_t$ is the total energy without the spin, $Z_c$ is the
number of nearest neighboring Fe ions, and $J_z$ is the exchange
integral. While the AFM total energy has the form,
\begin{equation}
E_{AFM}=E_t+Z_cJ_zS^2.
\end{equation}
Therefore, we can determine the exchange parameters from the energy
difference between different spin configurations. We set up the FM
structure with spin direction along $z$ axis while the G-type along
$x$ direction. From Fig. 3, it is obvious that $R3c$ space group
tends to produce a much favorable G-type structure. LSDA+U gives
lower AFM exchange integral than LSDA. It is worth pointing out
again that the anomaly takes place around the critical pressure
value. Exchange integral even lies in the same  level in these two
schemes. G-type structure has lower exchange interaction under the
whole pressure range. Two anomalies can be found at 9 and 12 GPa,
respectively. It is consistent with energy calculations. When
pressure exceeds the critical value, AFM exchange integral is
changing into a positive value where it applies to FM spin
configuration. This shows the AFM-FM transition occurs at the
critical pressure accompanying with the structural phase transition
which agrees well with the previous total energy calculations. The
phase graph under pressure is shown in Fig. 4. We suggest the
rhombohedral structure maintains before the critical pressure value
of 9 GPa. A combination of these three structures exist between 9
and 12 GPa, accompanying with an AFM-FM spin transition. A pure
orthorhombic phase will be found after 12 GPa while the FM spin
structure remains.

\subsection{G-type AFM vectors}
In G-type spin structure, we take into account the non-collinear and
SO coupling effect. The AFM vectors under pressure are reported in
Fig. 5. It is apparent that the AFM spin in $x$ direction cancel out
and a resultant magnetization along $y$ and $z$ direction can be
obtained due to the DMI.  DMI is caused by the interaction of
neighboring Fe sites which can be described by

\begin{equation}
E_{Fe1,Fe2}^{(2)}=\mathbf{J}_{Fe1,Fe2}^{(2)}(\mathbf{S_1}\cdot\mathbf{S_2})+
\mathbf{D}_{Fe1,Fe2}^{(2)}(\mathbf{S_1}\times\mathbf{S_2})+\mathbf{S}(R)\cdot
\Gamma_{Fe1,Fe2}^{(2)}\cdot \mathbf{S}_2,
\end{equation}
in the second order perturbation calculation\cite{25,26}. The first
term on the right hand side of the Eq. (6) corresponds to the usual
isotropic superexchange interaction, and the second term is the DMI.
The Hamiltonian for the system reads,

\begin{equation}
H_{BiFeO_3}=-2\sum_{<1i,2j>}\textbf{J}_{1i,2j}\mathbf{S}_{1i}\cdot\mathbf{S}_{2j}+\sum_{<1i,2j>}\textbf{D}_{1i,2j}\mathbf{S}_{1i}\times\mathbf{S}_{2j}.
\end{equation}
The first term is the symmetric superexchange, and the second one is
the antisymmetric DMI contribution. $\textbf{J}_{1i,2j}$  is a
constant similar to the exchange interaction. \textbf{D} is the DMI
constant  and determined by the sense of rotation of the neighboring
oxygen octahedra. \textbf{D} reads by the second order perturbation
in the case of one electron per ion
\begin{equation}
\textbf{D}_{Fe1,Fe2}^{(2)}=(4i/U)[b_{nn'}(Fe1-Fe2)C_{n'n}(Fe2-Fe1)-C_{nn'}(Fe1-Fe2)b_{n'n}(Fe2-Fe1)],
\end{equation}
where $U$ is the energy required to transfer one electron from one
site to its nearest neighbor, a parameter similar to on-site Coulomb
interaction in our $Ab$ $initio$ computation, and inversely
proportional to \textbf{D}. The spin value along $z$ direction is
depressed after pressure exceeds the critical value while spin
magnetization along $y$ direction has the opposite trend and
increases with pressure increasing. Firstly, the net magnetization
has  components along $z$ and $y$ direction simultaneously and
deviate away from the $z$ direction as pressure exceeds 12 GPa,
resulting in zero component along $z$, while that along $y$
direction increases and maitains a constant value after 12 GPa.  The
magnetization per unit cell is calculated based on the LSDA+U method
and it is underestimated in this way. Therefore greater value is
expected in LSDA calculations\cite{24}.
\subsection{Electronic properties}
In order to shed light on the electronic properties under pressure
unambiguously, the total density of states(DOS) before and after
exerting pressure are given in Fig. 6. The orbital resolved
DOS(ODOS) for Fe $3d$ orbitals are given in Fig. 7 and Fig. 8,
respectively. From Fig. 6, it can be seen that a semiconducting
band gap of 1.7 eV  is produced within LSDA+U  while the band gap
vanishes under pressure of 12 GPa, suggesting an obvious
insulator-metal (IM) transition. The IM transition is mainly caused
by the shift of the states of Fe $3d$ electrons in the vicinity of
fermi energy. The finite DOS of these electrons cut through the
fermi level and form a FM spin structure under 12 GPa.  From Fig. 7,
one can see all up-spin electrons are occupied and down-spin
electrons are in conduction band. While almost all up and down spin
states are partially filled under pressure in Fig. 8. Three-fold
degenerate states $t_{2g}$ coming from $d_{xy}$, $d_{yz}$, and
$d_{xz}$ orbitals remain degenerate, while $d_{z^2}$ orbital splits
from the two-fold degenerate states $e_g$. It is this orbital that
makes $e_g$-$e_g$ AFM interaction reduced. The splitting of the
orbital under pressure is significant, suggesting the complete
depression of AFM interaction and the occurrence of FM spin
structure. This is also the reason for the decrease of magnetization
of G-type AFM structure along $z$ direction under pressure.
Meanwhile the down-spin Fe $3d$ states under pressure cut across the
fermi energy and lead to the conducting behavior.

\section{ Conclusion}
The total energy, magnetic and electronic properties of BiFeO$_3$
under pressure are calculated based on the LSDA and LSDA+U scheme.
Results show two anomalies can be found at 9-10 GPa and 12 GPa,
respectively.  The first one is  the critical pressure for
first-order phase transition accompanying with AFM-FM transition.
Meanwhile the behavior under critical pressure also involves an IM
transition. The second one does not involve further magnetic
transition but structural transitions. The magnetization of the
G-type AFM spin structure in $y$ direction increases   while the $z$
direction component decreases, which can be explained by the
splitting of the $d_{z^2}$ orbital  from doubly degenerate $e_g$
states.



\clearpage

\begin{table}[!h]

\caption{ Current used lattice parameters for rhombohedral,
monoclinic, and orthorhombic structures}

\begin{center}

\tabcolsep=8pt
\begin{tabular}{@{}cccc}
\hline\hline
  &$R3c$&$Cm$   &$Pnma$\\
\hline
a({\AA})& 5.459& 5.7900 & 5.5849 \\
b({\AA})& &5.6899  &7.6597 \\
c({\AA})& &  4.1739& 5.3497\\
$\alpha$(\textordmasculine )&60.36& &  \\
$\beta$(\textordmasculine )& &91.99& \\
V({\AA$^3$}) & 115.98 & 137.42  & 113.12   \\
Bi& (2a):0,0,0& (2a):0.9376,0,0.0685 & (4c):0.0536,0.2500,0.9886  \\
Fe&(2a):0.2308,0.2308,0.2308& (2a):0.5110,0,0.4961& (4b):0,0,0.5  \\
O&(6b):0.5423,0.9428,0.3980& (2a):0.5626,0,0,9489  & (4c):0.9750,0.2500,0.4060;\\
&& (4b):0.7958,0.7603,0.4231&(8d):0.2000,0.9540,0.1945 \\
\hline\hline
       \end{tabular}

       \end{center}
       \end{table}

\clearpage

\raggedright \textbf{Figure captions:}

Fig.1 FM magnetization for three phases of BiFeO$_3$ under pressure.

 Fig.2 The total energy as functions of pressures.

 Fig.3 The exchange integrals for different spin structures as functions of pressure.

 Fig. 4 The phase graph of BiFeO$_3$ under pressure.

 Fig. 5 G-type AFM vectors variations with respect to pressure.

 Fig. 6 Total DOS under ambient and transition pressure.

 Fig. 7 ODOS for Fe $d_{xy},d_{yz},d_{z^2},d_{xz}$, and
$d_{x^2-y^2}$ orbitals under ambient pressure.

 Fig. 8 ODOS for Fe $d_{xy},d_{yz},d_{z^2},d_{xz}$, and
$d_{x^2-y^2}$ orbitals under transition pressure.

\clearpage
\begin{figure}
\centering
\includegraphics{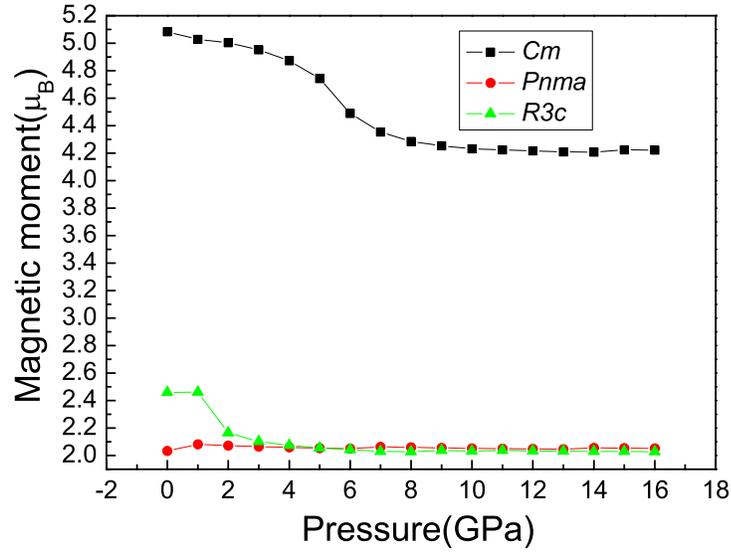}
\caption{FM magnetization for three phases of BiFeO$_3$ under
pressure.}
\end{figure}

\begin{figure}
\centering
\includegraphics{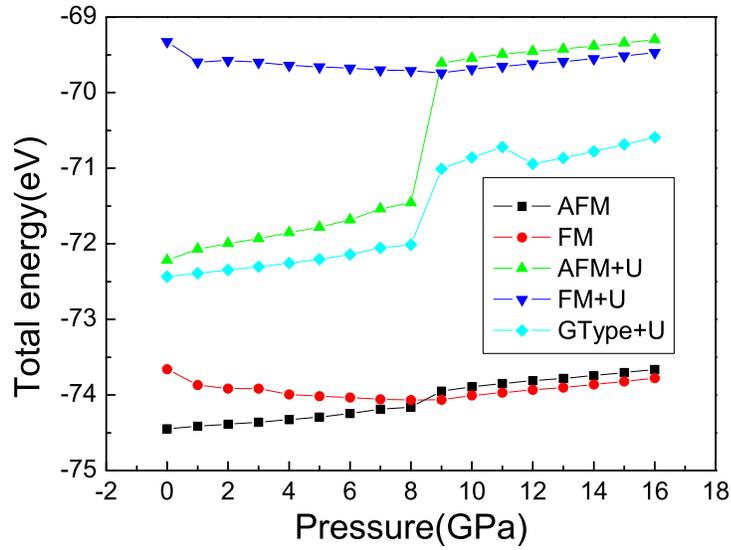}
\caption{ The total energy as functions of pressures.}
\end{figure}

\begin{figure}
\centering
\includegraphics[width=8cm]{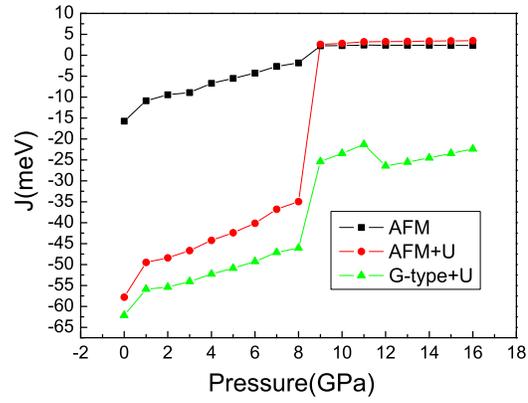}
\caption{The exchange integrals for different spin structures as
functions of pressure.}
\end{figure}

\begin{figure}
\centering
\includegraphics{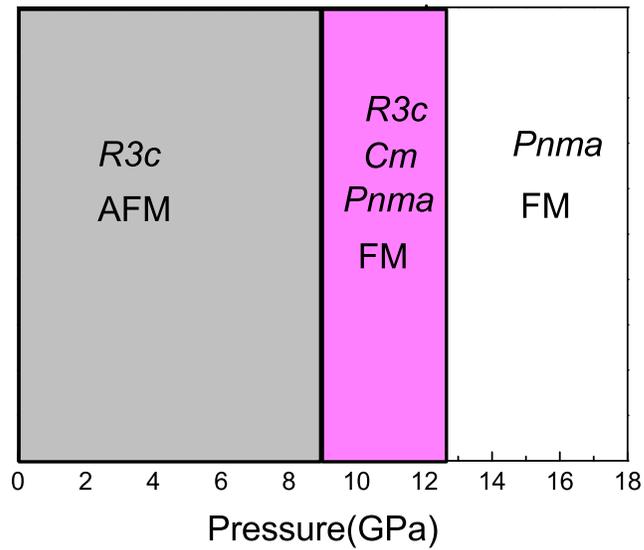}
\caption{The phase graph of BiFeO$_3$ under pressure.}
\end{figure}

\begin{figure}
\centering
\includegraphics{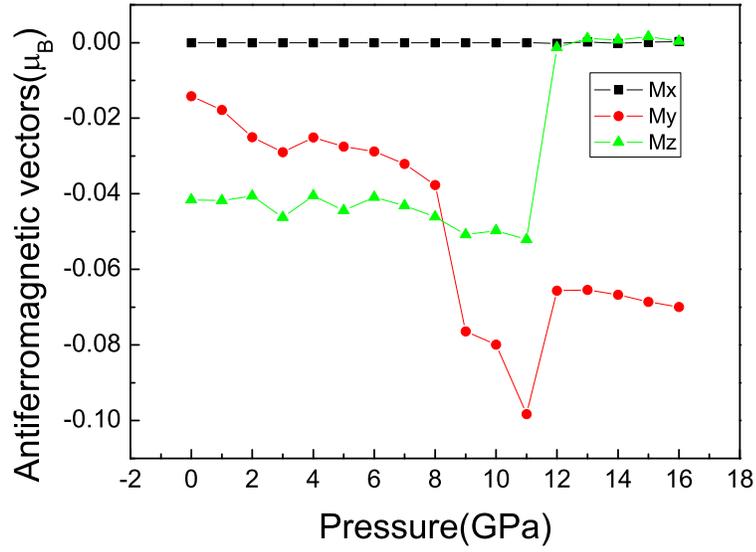}
\caption{G-type AFM vectors variations with respect to pressure.}
\end{figure}

\begin{figure}
\centering
\includegraphics{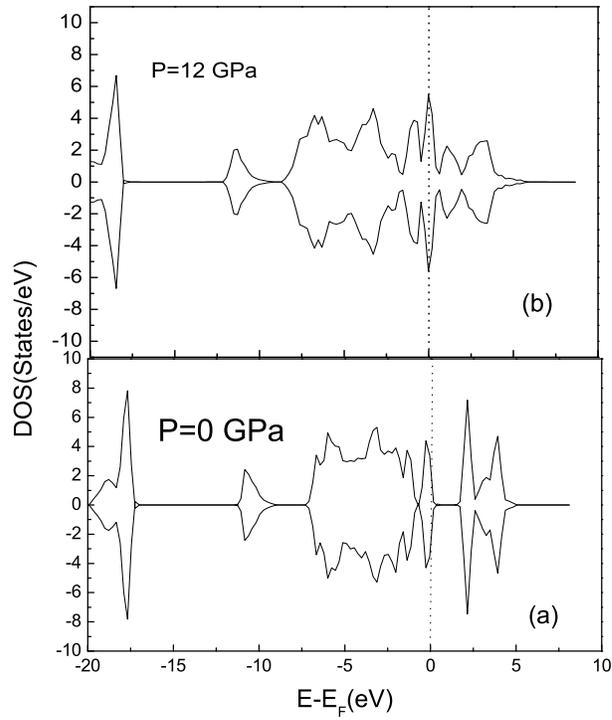}
\caption{Total DOS under ambient and transition pressure.}
\end{figure}

\begin{figure}
\centering
\includegraphics{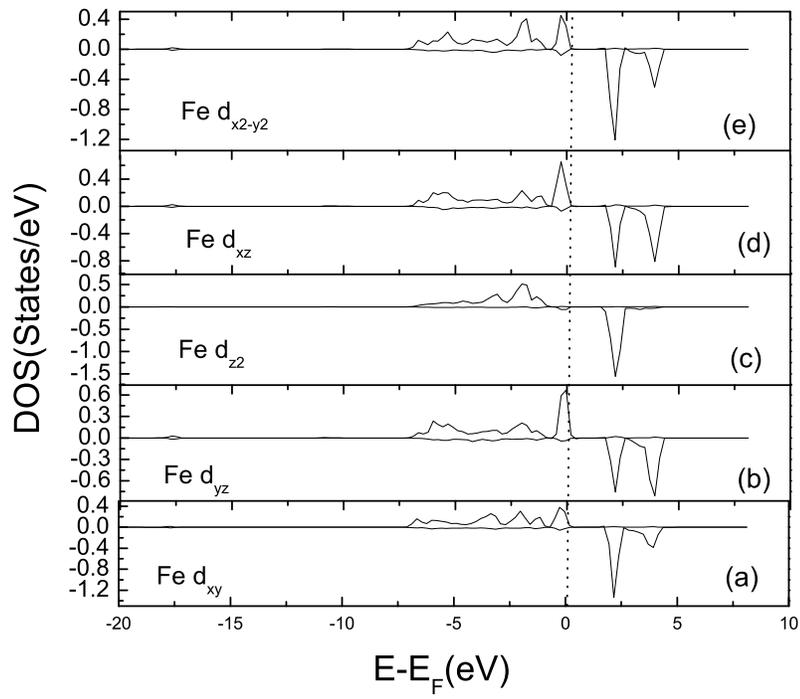}
\caption{ODOS for Fe $d_{xy},d_{yz},d_{z^2},d_{xz}$, and
$d_{x^2-y^2}$ orbitals under ambient pressure.}
\end{figure}

\begin{figure}
\centering
\includegraphics{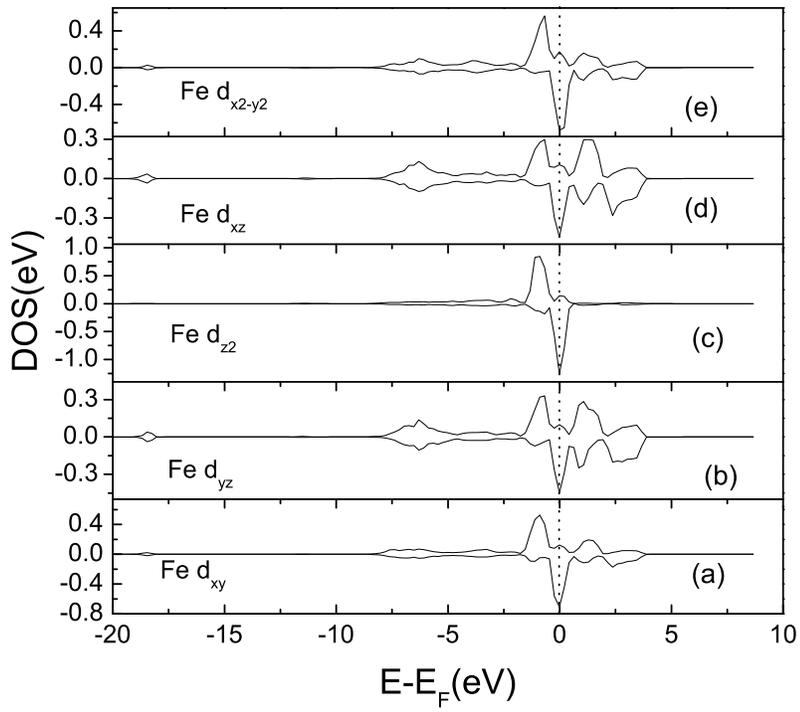}
\caption{ODOS for Fe $d_{xy},d_{yz},d_{z^2},d_{xz}$, and
$d_{x^2-y^2}$ orbitals under transition pressure.}
\end{figure}

\end{document}